\documentclass[conference]{IEEEtran}
\usepackage{ifpdf}

\ifCLASSINFOpdf
  \usepackage[pdftex]{graphicx}
\else
  \usepackage[dvips]{graphicx}
\fi
\usepackage[cmex10]{amsmath}
\usepackage{url}

\usepackage{color}
\usepackage[USenglish]{babel}
\usepackage[loose]{subfigure}
\usepackage{multirow}
\usepackage{rotating}
\usepackage{listings}
\usepackage[htt]{hyphenat}
\usepackage[small,bf]{caption}

\newcommand{\lineref}[1]{\hyperref[#1]{line~\ref*{#1}}}
\newcommand{\algref}[1]{\hyperref[#1]{Algorithm~\ref*{#1}}}

\DeclareMathAlphabet{\mathpzc}{OT1}{pzc}{m}{it}

\newcommand{\blaze}{{\it Blaze}}

\DeclareMathAlphabet{\mathpzc}{OT1}{pzc}{m}{it}
\newcommand{\ublas}{{\it uBLAS}}

\DeclareMathAlphabet{\mathpzc}{OT1}{pzc}{m}{it}
\newcommand{\blitz}{{\it Blitz++}}

\DeclareMathAlphabet{\mathpzc}{OT1}{pzc}{m}{it}
\newcommand{\mtl}{{\it MTL4}}

\DeclareMathAlphabet{\mathpzc}{OT1}{pzc}{m}{it}
\newcommand{\eigen}{{\it Eigen3}}

\DeclareMathAlphabet{\mathpzc}{OT1}{pzc}{m}{it}
\newcommand{\armadillo}{{\it Armadillo}}

\DeclareMathAlphabet{\mathpzc}{OT1}{pzc}{m}{it}
\newcommand{\gmm}{{\it GMM++}}

\graphicspath{{pics/}}

\definecolor{gray}{rgb}{0.95,0.95,0.95}
\definecolor{lightgray}{gray}{0.45}
\definecolor{myblue}{rgb}{0.22,0.22,0.61}
\definecolor{myred}{rgb}{0.6,0,0}

\begin{document}


\title{Model-guided Performance Analysis of the Sparse Matrix-Matrix Multiplication}




\author{\IEEEauthorblockN{Tobias Scharpff\IEEEauthorrefmark{1},
Klaus Iglberger\IEEEauthorrefmark{1},
Georg Hager\IEEEauthorrefmark{2} and
Ulrich R{\"u}de\IEEEauthorrefmark{1}}
\IEEEauthorblockA{\IEEEauthorrefmark{1}Chair for System Simulation,
University Erlangen-Nuremberg,
Erlangen, Germany\\ Email: \{tobias.scharpff,klaus.iglberger,uli.ruede\}@fau.de}
\IEEEauthorblockA{\IEEEauthorrefmark{2}Erlangen Regional Computing Center,
University Erlangen-Nuremberg,
Erlangen, Germany\\ Email: georg.hager@fau.de}}


\maketitle


\begin{abstract}
  Achieving high efficiency with numerical kernels for sparse matrices
  is of utmost importance, since they are part of many simulation
  codes and tend to use most of the available compute time and
  resources. In addition, especially in large scale simulation
  frameworks the readability and ease of use of mathematical
  expressions are essential components for the continuous maintenance,
  modification, and extension of software.

  In this context, the sparse matrix-matrix multiplication is of special
  interest. In this paper we thoroughly analyze the single-core performance
  of sparse matrix-matrix multiplication kernels in the \blaze{} Smart
  Expression Template (SET) framework. We develop simple models for estimating
  the achievable maximum performance, and use them to assess the
  efficiency of our implementations.  Additionally, we compare these
  kernels with several commonly used SET-based 
  C++ libraries, which, just as \blaze{}, aim at combining the
  requirements of high performance with an elegant user interface.

  For the different sparse matrix structures considered here, we show 
  that our implementations are competitive or faster than those of
  the other SET libraries for most problem sizes on a current 
  Intel multicore processor.  
\end{abstract}


%
\IEEEpeerreviewmaketitle


\section{Motivation}
\label{sec:motivation}

Various popular simulation algorithms in high performance computing (HPC), 
such as computational dynamics for
rigid bodies, rely on sparse matrix-matrix multiplication (spMMM)
as one of their computational kernels. 
Due to its central role in the applications and its
computational complexity it is of vital importance to have highly optimized
implementations. However, apart from performance, other metrics such as
programmability, readability and foremost maintainability are crucial
for a successful long-term software development effort. Yet these metrics usually
play only a minor role in HPC software development. Although
there exist several approaches to provide fast spMMM implementations,
these libraries, like most HPC software efforts, strictly focus on high
performance but do not so well in most other software metrics (see
also~\cite{iglberger:2012:SISC}). This neglect especially endangers complex,
long-term software developments due to impeded maintenance work. However,
the maintainability of software should be of major interest: on average, 60\%
of the total software development costs is spent in maintenance, where
long-term projects usually lean towards higher maintenance costs~\cite{davis:09}.
Improving the maintainability immediately leads to less effort in software
modification and extension and subsequently to fewer software defects.

This realization is the driving force behind several C++ Smart Expression
Template (SET) math libraries. These libraries attempt to combine highly
optimized math kernels for vector and matrix operations with the
advantages of a domain-specific language. They include an intuitive
formulation of mathematical operations, high readability, and easy
modification of operations (see for instance Listing~\ref{lst:spMMM_in_blaze}).

\begin{lstlisting}[numbers=left,
                   frame=tb,
                   caption={spMMM formulation in the Blaze SET math library.},
                   label={lst:spMMM_in_blaze},]
blaze::CompressedMatrix<double,rowMajor> A, B, C;
// ... Initialization of matrices A and B
C = A * B;
\end{lstlisting}

In this paper we focus on the optimization of the sequential spMMM
algorithm in the \blaze\ SET library, and compare the resulting
performance characteristics for two chosen sparse matrices with
similar high-performance SET-based frameworks. It will be clear that
such a comparison can only make sense when the analysis is performed
over a wide range of problem sizes, which rules out an extensive
survey of popular matrix collections. We recognize that such a survey
would be desirable. In this work we prefer a deeper analysis with
more insight, however, and leave the survey to future research.

This paper
is organized as follows. In Section~\ref{sec:related_work}
we give a short overview of other C++ math libraries that follow an
approach similar to \blaze, before Section~\ref{sec:benchmark_platform} briefly
summarizes the details of our benchmark platform and benchmarking
strategy. In Section~\ref{sec:spMMM} we describe basic tasks and
necessary steps for spMMM, together with appropriate performance
models. Here we also demonstrate the general suitability of the SET
methodology for HPC in terms of performance and the advantages in
terms of software development. The optimized kernels are benchmarked
and compared to several other SET-based C++ math libraries in
Section~\ref{sec:results}. Section~\ref{sec:conclusion} concludes the
paper and provides suggestions for future work.


\section{Related Work}
\label{sec:related_work}

The C++ programming language provides the feature to directly overload mathematical operators,
which enables a very intuitive application of mathematical operations. However, due to the
necessary creation of a temporary in each operation, the performance of classic C++
operator overloading cannot compete with other approaches. A reputed solution are Expression
Templates (ET), which due to lazy evaluation of the result promise optimized performance.
The first ET-based C++ library for dense arithmetic was \blitz{}~\cite{blitz}. This framework, written
by the inventor of ETs, Todd Veldhuizen, has been recognized as a pioneer in the area of C++
template metaprogramming~\cite{Abrahams:2005:C++TMP}. The Boost \ublas{}
library~\cite{boost_ublas} is one of the most widespread ET math libraries since it is distributed
together with the Boost library~\cite{boost}. In contrast to \blitz{} it additionally provides
sparse matrices and vectors. However, in~\cite{iglberger:2012:SISC} we have demonstrated that
the assumption that ETs are a performance optimization is not justified, and have introduced an
improved solution by the Smart Expression Template (SET) methodology. Among other features,
SETs encapsulate performance-optimized compute kernels like those provided by the BLAS
and LAPACK standards. An early example for a SET library is \armadillo{}~\cite{armadillo},
which is restricted to dense linear algebra operations, but employs SETs to integrate BLAS and LAPACK 
for optimized performance. The same feature is provided by \mtl, which additionally includes
sparse matrix operations. An alternative with similar functionality is the \gmm{} library~\cite{gmm},
which allows to use ATLAS~\cite{atlas} as BLAS backend. Numerics involving dense and sparse matrices
and vectors, the use of optimized kernels, and the advanced SET features of intrinsics-based
vectorization of non-BLAS operations and automatic expression optimization are supported by
the \eigen{}~\cite{eigen} and the \blaze{}~\cite{blaze} libraries. In contrast to \eigen, which
provides optimized kernels for all basic operations, \blaze{} uses BLAS subroutines for BLAS
level 2 and 3 operations.

In~\cite{iglberger:2012:SISC} we have analyzed several of these ET implementations in detail 
and have introduced the notion of SETs and our SET library \blaze{} in particular.
In~\cite{iglberger:2012:HPCS12} we have extended our analysis to more ET-based libraries,
focused on the optimization and vectorization capabilities for dense arithmetic, and presented
performance results for the CG algorithm, which is fundamental for many applications. In this
work we expand our analysis to sparse arithmetic and the sparse matrix-matrix multiplication
(spMMM) in particular.

Much work has been devoted to sparse matrix based algorithms and
efficient implementations in the past. However, most publications deal
with parallel sparse matrix-vector multiplication
\cite{Bisseling95sparsematrix,Williams09,schubert:ppl11}, since it is
of pivotal importance in solving sparse linear systems and sparse
eigenvalue problems. While there has also been substantial work on
sparse matrix-matrix multiplication, it mostly deals with execution
and communication efficiency in the parallel case
\cite{mccoll99,BerkeleyspMMM13}. Here, however, we only cover the
sequential kernel any try to understand its features in a well-defined
setting, and especially in the context of SET frameworks. Consequently,
issues of load and communication balancing, which would be crucial
in the parallel case, do not arise.

\section{Benchmark Platform and Test-Cases}
\label{sec:benchmark_platform}
An Intel Sandy Bridge i7-2600 CPU was used for all benchmarks. Using only one of the four
cores it runs at 3.8~GHz with 8~MB of shared L3 cache, 256~kB of L2 cache 
and 32~kB of L1 data cache. The maximum
achievable main memory bandwidth (as measured by the STREAM benchmark~\cite{mccalpin:2007}) is
about 18.5~GB/s. For each non-zero element of a sparse matrix we store the value as double precision
floating point number and an index as a 64-bit integral value. Since we concentrate
on general sparse matrices and there is no vector gather instruction in current x86 designs, 
we do not utilize SIMD vectorization but run scalar code.
This means that the CPU is
capable of performing one double precision floating point multiplication and one double precision
floating point addition as well as either two load or one load~(LD) and one store~(ST) instruction
per cycle~\cite{ia32opt:2012}. Therefore the theoretical peak performance is 7.6~GFlop/s.

The benchmark platform runs an openSuse~12.1 and we use the GNU~g++ compiler with
the following compiler flags: \texttt{-Wall -Werror -ansi -pedantic -O3 -mavx -DNDEBUG}

We use the Blazemark benchmark suite, which ships with \blaze{}, for a direct comparison
of the different libraries. It uses the same random seed for all libraries and care is taken that
randomly generated numbers and structures are identical for all tested libraries. We extended
the Blazemark to have the option to compare not only different libraries but also multiple
implementations of the \blaze{} spMMM kernels. To make sure that all measured times are
accurate the Blazemark runs short test-cases several times until the total runtime exceeds
two seconds. Furthermore, each test is performed at least 5 times and the best result is taken
as the final measurement. The number of floating point operations per second (Flops/s) for the
spMMM are calculated as follows: The number of required multiplications is
\begin{equation*}
\sum^{n-1}_{k=0} \bar a_{k}* \bar b_{k},
\end{equation*}
where $\bar a_{k}$ is the number of non-zeros in the $k$-th column of $A$, and $\bar b_{k}$ is
the number of non-zeros in the $k$-th row of $B$. The number of additions required is always
bounded by the required number of multiplications. We always use the worst case assumption
to calculate
the Flops, which means that the overall number of floating point operations is approximately 
twice the number of multiplications.

Two different input matrices over a range of problem sizes are used to review the outcome of our
analysis. The first test-case multiplies two five-band matrices, which are created by using
a 5-point stencil resulting from a finite difference discretization of a Dirichlet boundary
value problem on a square.
All graphs showing the result of multiplying two of these five-band matrices are marked
with (FD). The second test-case uses two randomly generated quadratic matrices. For each
matrix five random numbers are placed on random locations in each row. This allows for a
good comparability between the two test-cases in terms of the fill rate of the matrices.
Whenever a graph shows the outcome of the multiplication of two randomly generated
matrices it is marked with (random).

\section{Implementation and performance analysis of the spMMM kernel}
\label{sec:spMMM}
For a thorough analysis it has turned out to be convenient to split the spMMM kernel in two 
logically independent parts:
The pure computation and the actual storing of the results. 

\subsection{The pure spMMM computation kernel}
\label{sec:computation}
Looking only at the pure computation of the spMMM allows to implement this part of the
kernel and be sure that it works at the highest possible performance without any interference of
additional data accesses for storing the result. 
In \blaze{} we use implementations of the two well known formats ``compressed sparse
row'' (CSR) and ``compressed sparse column'' (CSC)~\cite{Barret:1994}. These formats
usually show good performance for general matrices on general-purpose cache-based 
microprocessors.
Both formats use an array of pointers, which provides an immediate access
to a specific row (CSR) or column (CSC).

The classic way of calculating a matrix-matrix product \mbox{$C = A * B$} is to perform a dot
product-like operation between a row of $A$ and a column of $B$ for each element in the resulting
matrix. To achieve optimal performance with this approach, the format should be CSR for matrix 
$A$  and CSC for $B$, while the format of $C$ is irrelevant. The problem is that both vectors are
sparse and therefore the operation suffers from all known issues of sparse vector-vector
multiplications. Furthermore the results of these ``dot products'' are zero most of the time.
Optimizations such as unrolling or blocking, which would lead to increased computational
intensity~\cite{Hager:2011:HPCSE,Williams09}, 
rely on exploiting specific matrix structures and will not be explored here.

\begin{figure}[h]
   \centering
   \includegraphics*[width=0.48\textwidth]{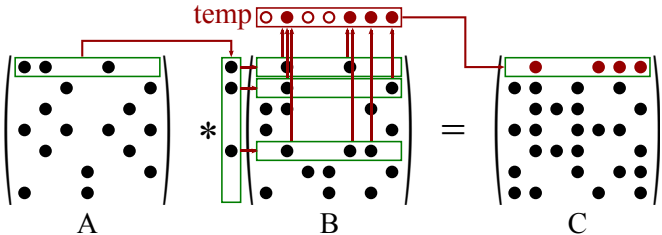}
   \caption[Sketch of a spMMM with the row-major algorithm]{
      \label{fig:zeilenorientiert}
      Sketch of a spMMM with the row-major algorithm.
   }
\end{figure}

Another algorithm, optimized for a set of three CSR or three CSC matrices, was introduced
in~\cite{gustavson:1978}. As shown in Figure~\ref{fig:zeilenorientiert}
it multiplies each non-zero value $a_{r,c}$ of row $r$ of matrix $A$ with all non-zero entries
$b_{c,x}$ of matrix $B$. The intermediate results are collected in a dense temporary vector, which
is initially filled with zeros, by just adding each result to the current value at the position $x$
of the temporary vector. If this is done for all non-zero values of row $r$ of matrix $A$, the
vector is a dense representation of the $r$th row of the resulting matrix. Note that the approach
can also be applied to column-major matrices in the spMMM with three CSC matrices.

In case one of the two matrices is available in CSR format and the other in CSC format it
turns out to be more efficient to convert one of the matrices to the other format instead of
providing a fallback to the ``classic'' algorithm. The effort to convert the format is linear
in the number of non-zero entries. Therefore it is not necessary to implement a total of eight
computation kernels for all eight possible combinations. It is sufficient to convert one of
the two matrices to be able to use the row-major or column-major algorithm.

In order to arrive at a realistic upper performance limit for our computational kernels we employ
a simple bandwidth-based performance model~\cite{Williams:EECS-2008-134,Hager:2011:HPCSE}: 
The maximum performance for a loop is
$$
P = \max\left( P_\mathrm{max}, \frac{b_{\mathrm{max}}}{B_\mathrm c}\right)~,
$$
where $b_\mathrm{max}$ is the bandwidth of the relevant data path
in bytes/s and $B_\mathrm c$ is the loop's code balance:
$$
B_{\mathrm{c}}
= \frac{\textrm{Data traffic [Bytes]}}{\textrm{Floating point operations [Flops]}}
$$
This model works well if the performance of the loop is dominated by the
data transfers to and from a single data path. Other effects, such as
dependencies, abundant branch mispredictions, or the general inability of 
a single core to saturate the bandwidth of some memory hierarchy levels,
may cause significant deviations from the model \cite{Hager:CPE12}. However,
it is still a valuable starting point for a loop-based performance analysis,
since it provides a ``light speed'' estimate. We concentrate here on modeling
the more advanced implementations of the spMMM kernel, since the naive 
version is plagued by conditional branches and erratic access patterns,
which are not easily modeled.


\begin{lstlisting}[numbers=left,
                   frame=tb,
                   caption={The row-major computation kernel without storing the result to the matrix $C$.},
                   label={lst:Zeilenorientiert},]
void compute(
   CSRMatrix& C,
   const CSRMatrix& A,
   const CSRMatrix& B )
{
  typedef CSRMatrix::const_iterator  iterator;

  // Estimate the number of elements in matrix C
  nnzEstimation( C, A, B );

  // Temporary vector to store the result row-wise
  std::vector<double> temp( C.columns(), 0.0 );

  // Loop over all rows of the target matrix
  for( std::size_t cy = 0; cy < C.rows(); ++cy )
  {
    iterator ait( A.begin(cy) );
    iterator const aend( A.end(cy) );
	
    // Loop over the non-zero entries of the
    //   current row of A
    for( ; ait!=aend; ++ait )
    {
      std::size_t const indexA( ait->index() );
      double const      valueA( ait->value() );
      
      iterator bit( B.begin(indexA) );
      iterator const bend( B.end(indexA) );

      // Loop over the non-zero entries of the
      //   current row of B
      for( ; bit!=bend; ++bit ) @\label{lst:Zeile_innen_beginn}@
      {
        size_t const indexB( bit->index() ); // LD
		
        // Update value
        // LD + Mult + LD + ADD + ST
        temp[indexB] += valueA * bit->value(); @\label{lst:Zeile_rechnen}@
      } @\label{lst:Zeile_innen_end}@
    }

    // Write result to matrix C
    // Reset all entries of vector temp to 0.0
  }
}
\end{lstlisting}

Listing~\ref{lst:Zeilenorientiert} shows the code for the row-major computation kernel.
The inner loop between lines~\ref{lst:Zeile_innen_beginn} and \ref{lst:Zeile_innen_end}
has a code balance of 16~Bytes/Flop. We assume that the update to the \verb.temp[]. vector
causes a load and a store to the relevant memory hierarchy level, but ignore non-consecutive
accesses, which would lead to excess data traffic. Hence, the predictions of the
balance model must be seen as best-case values. Within the L1 cache
this leads to a maximum theoretical performance of 3800~MFlops/sec at 3.8~GHz clock
frequency, whereas in memory the limit is 1140~MFlops/sec. 

\begin{figure}[h]
   \centering
   \includegraphics*[width=0.48\textwidth]{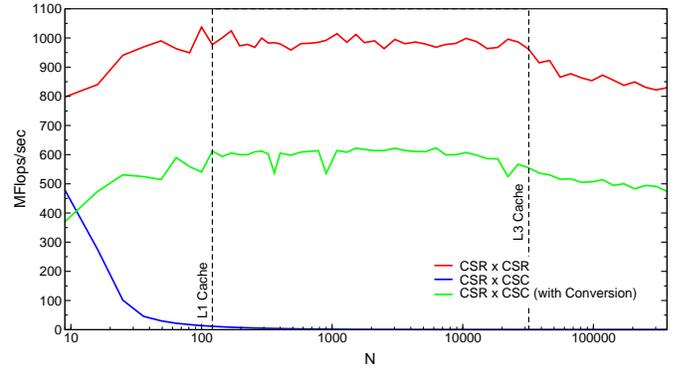}
   \caption[Performance results of the pure computation (FD)]{
      \label{fig:nurrechnen_fd}
      Performance results of the pure computation (FD). The maximum theoretical
      performance beyond the L3 limit is 1140\,MFlops/sec.
   }
\end{figure}

Figure~\ref{fig:nurrechnen_fd} shows performance results versus problem size (number of matrix
rows) for the 5-point 
finite difference stencil matrices. The row-major algorithm (\mbox{CSR $\times$ CSR}) 
clearly achieves the best results for \mbox{CSR $\times$ CSR} and even comes 
close to the theoretical performance of 1140~MFlops/sec beyond the L3 cache limit. 
Even if the right-hand side operand is given as a CSC matrix and is therefore internally converted to 
CSR (\mbox{CSR $\times$ CSC} (with conversion)), still about 50\% of the original performance is 
achieved. The classic \mbox{CSR $\times$ CSC} kernel cannot compete with the the row-major approach 
due to the problems mentioned before. The fact that the row-major algorithm's performance only drops 
slightly for matrices that do not fit into the L3 cache anymore
shows that the balance model is problematic for in-cache situations, and more advanced
modeling techniques would be required~\cite{Hager:CPE12}. 
All data of the left-hand side matrix is traversed 
with stride one. For the right-hand side operand the prefetcher can easily 
predict which data to load, thanks to the fixed five-band pattern of the matrix.  

\begin{figure}[h]
   \centering
   \includegraphics*[width=0.48\textwidth]{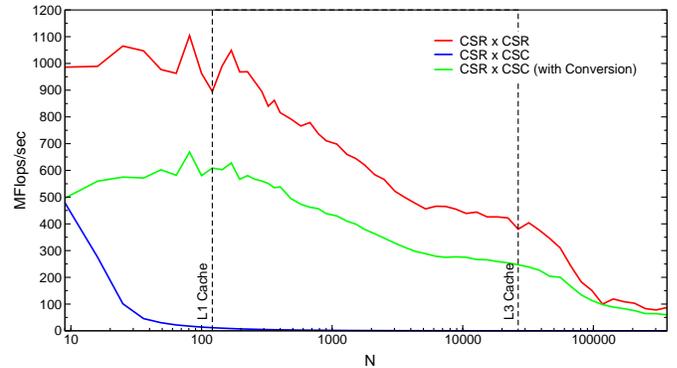}
   \caption[Performance results of the pure computation (random)]{
      \label{fig:nurrechnen_random}
      Performance results of the pure computation (random). The maximum theoretical
      performance beyond the L3 limit is 1140\,MFlops/sec.
   }
\end{figure}

Figure~\ref{fig:nurrechnen_random} shows the results for the test case which uses randomly generated
spares matrices. The classic \mbox{CSR $\times$ CSC} algorithm is not influenced by the structure
of the matrices and therefore shows the same bad performance we saw in 
Figure~\ref{fig:nurrechnen_fd}. The row-major approach clearly achieves better results. However,
because of the random structure of the left-hand side operand the prefetcher does not work
optimally for the right-hand side matrix; thus, performance goes down with growing
problem sizes. The classic approach does not show any significant performance for problem sizes 
greater than $N=200$. Compared to this the row-major approach shows a much better performance even 
for huge matrices that do not fit into the L3 cache anymore, and also if the right-hand side
matrix hast to be converted to the other format.  Due to the cache-unfriendly
access patterns the calculated performance limits cannot nearly be reached for this matrix.

Note that the general guideline to have a regular matrix structure for best performance
is valid predominantly in view of the left-hand side matrix $A$; the performance is largely
independent of the structure of $B$.

\subsection{Storing the spMMM result}
\label{sec:storing}


The algorithm in Listing~\ref{lst:Zeilenorientiert} only calculates all the entries for the result matrix, but
never actually stores them to the matrix object. Therefore all further optimization is driven by the requirement
to access the memory when storing the result in the most efficient way.

In this context, estimating the number of non-zero entries in the resulting matrix is an
essential aspect. It is of highest importance to prevent frequent dynamic memory
allocations during the calculation. Therefore an estimation of the final number of
non-zero entries is required that never underestimates and, if possible, only slightly
overestimates the needed memory. We found that the number of multiplications
required to perform the spMMM (see~\ref{sec:benchmark_platform})
is a good estimate. Each intermediate result either takes a place which is still zero or is
added to another intermediate result. Due to this fact the number is always equal or higher
than the number of non-zeros in the resulting matrix. Using this estimation the memory
allocation is only done once at the beginning of the kernel.

Another performance-critical part is the interface for storing the values in the resulting
matrix. Our implementation of the CSR/CSC formats provides two low-level functions
for this. First the \texttt{append} function, which appends an entry. 
It is the programmer's responsibility to append values in increasing row order and,
within each row, in increasing column order. The second function is \texttt{finalize}, which
marks the end of a row after all values have been appended. It has to be called after
each row and leaves the matrix in a consistent state  (note
that the CSC format is handled accordingly). Streaming the 
results in this way has the advantage that all the values are stored in one successive memory block,
and the underlying data structure for the row access is only modified once per spMMM.

\begin{figure}[h]
   \centering
   \includegraphics*[width=0.48\textwidth]{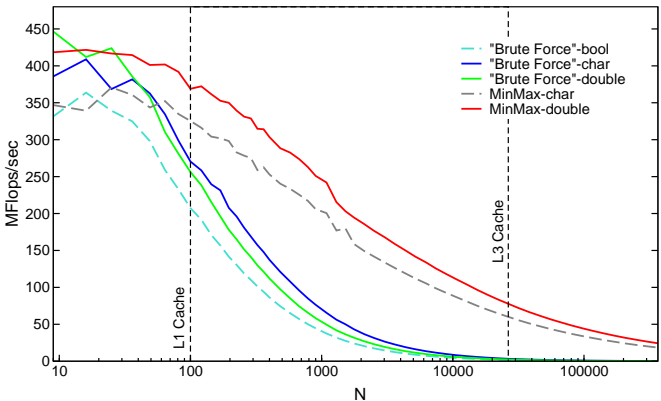}
   \caption[Comparison of different ``brute force'' and ``min-max'' kernels (FD)]{
      \label{fig:bf_minmax_fd}
      Comparison of different ``Brute Force'' and ``MinMax'' kernels (FD) for the 
      complete spMMM.
   }
\end{figure}

\begin{figure}[h]
   \centering
   \includegraphics*[width=0.48\textwidth]{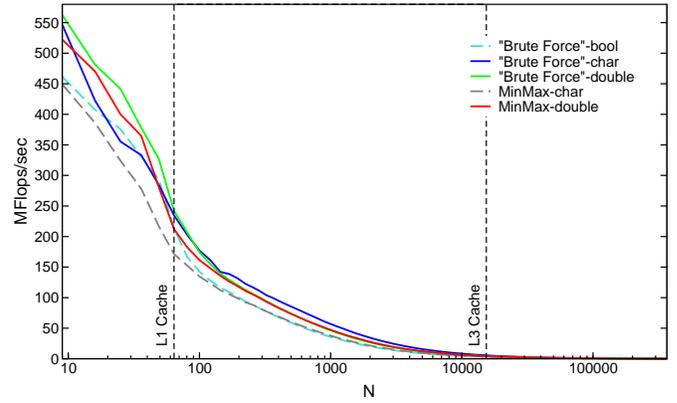}
   \caption[Comparison of different ``brute force'' and ``min-max'' kernels (random)]{
      \label{fig:bf_minmax_fd_random}
      Comparison of different ``Brute Force'' and ``MinMax'' kernels (random) for the 
      complete spMMM.
   }
\end{figure}

We have shown above that the row-major algorithm (see Listing~\ref{lst:Zeilenorientiert}) is very efficient.
It calculates a dense representation of each result row, which subsequently has to
be stored in the sparse result matrix. However, the way the temporary vector is
converted to a sparse row is crucial. A first alternative is a brute force approach,
which iterates over the \verb.double. values of the temporary vector and appends all non-zero
values to the resulting matrix (``Brute Force''-\verb.double.). To reduce the amount of memory
that has to be traversed the second approach is to use an additional lookup vector,
either of type \texttt{bool} (``Brute Force''-\verb.bool.) or \texttt{char} (``Brute Force''-\verb.char.).
In the STL a \texttt{std::vector<bool>} is implemented as a bit field~\cite{meyers:2008:ESTL}
and can therefore hold information for 512 positions per cache line instead of 8
\texttt{double}s or 64 \texttt{char}s. Figure~\ref{fig:bf_minmax_fd} shows
the performance results for the \mbox{CSR $\times$ CSR} ``brute force'' kernels
for the 5-point finite difference stencils and Figure~\ref{fig:bf_minmax_fd_random}
shows the corresponding results for the randomly created matrices.

Despite the fact that ``Brute Force''-\verb.bool. accesses the least memory it has to perform
additional Boolean operations for each entry, which leads to the worst performance in both
cases. Also in both cases the additional \texttt{char} vector increases the performance
slightly compared with the ``Brute Force''-\verb.double. approach without die additional
lookup vector. Also shown are our ``MinMax'' kernels, which basically do the same as
the ``Brute Force'' kernels, but additionally keep track of the lowest and highest index
of the non-zero entries in the temporary vector. Especially in the test-case with the
five-band matrices this optimization gives a considerable performance boost. Notably,
using the additional \texttt{char} vector hurts the performance of ``MinMax'' considerably. 
With the ``MinMax'' kernel each checked entry of temporary vector is more likely a non-zero 
value and therefore the advantage of the lookup vector is not big enough to
compensate the extra effort.

Even though the ``MinMax'' approach is better than ``Brute Force,'' both influence
the performance significantly. In addition, the bigger the problem sizes the more the performance
suffers compared to the pure computation kernel. With the problem size also the length
of the temporary vector and the number of elements in the minimum-maximum range
increases, but the absolute number of non-zeros does not change significantly. 

\begin{figure}[h]
   \centering
   \includegraphics*[width=0.48\textwidth]{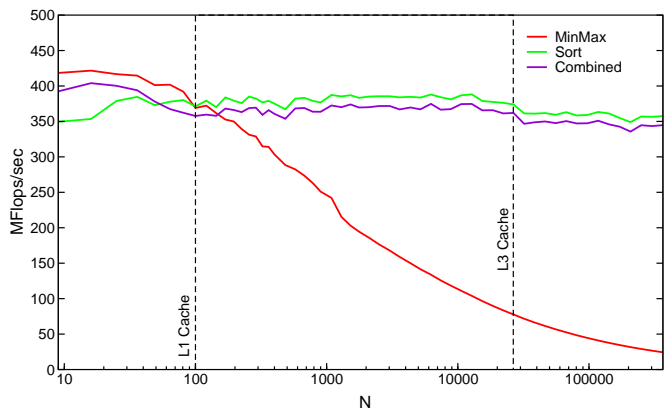}
   \caption[Comparison of the ``MinMax'' approach with the ``Sort'' approach (FD)]{
      \label{fig:minmax_stdsort_fd}
      Comparison of the ``MinMax'' approach with the ``Sort'' approach (FD) for the 
      complete spMMM.
   }
\end{figure}

\begin{figure}[h]
   \centering
   \includegraphics*[width=0.48\textwidth]{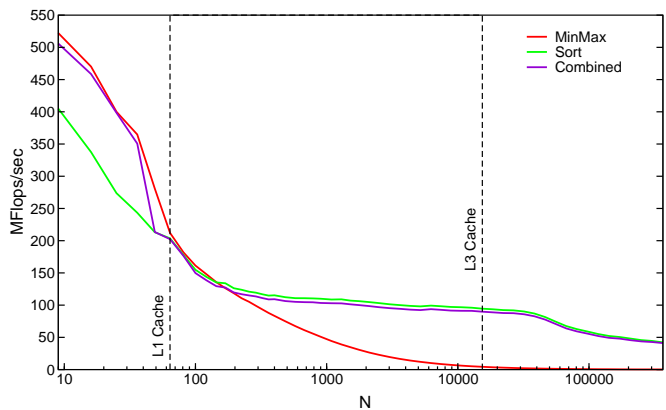}
   \caption[Comparison of the ``MinMax'' approach with the ``Sort'' approach (random)]{
      \label{fig:minmax_stdsort_random}
      Comparison of the ``MinMax'' approach with the ``Sort'' approach (random) for the 
      complete spMMM.
   }
\end{figure}

The next approach is to store all indices for non-zero elements within a row in a 
separate vector, which is usually small enough to fit into any cache level. 
After the complete row is calculated the
few entries of the vector that hold the indices are sorted using \texttt{std::sort}, and then only these positions of
the temporary vector are appended to the resulting matrix. Figure~\ref{fig:minmax_stdsort_fd}
shows the performance results for the \mbox{CSR $\times$ CSR} for the five-point finite difference stencils
with the sorting kernel (Sort) and Figure~\ref{fig:minmax_stdsort_random} illustrates the  corresponding
results for the test-case with the randomly generated matrices.
It shows that the performance drawback of the
sorting approach does not significantly increase with the problem size.

For both test-cases the ``MinMax'' approach still performs better at small problem
sizes. Hence, the final approach is to combine the ``MinMax'' and ``Sort'' kernels to the
new ``Combined'' kernel. The decision which of the two storing strategies to use is
performed for every single row. Note that it is more important that the decision can be done
quickly than that it is precise, as it is performed for every row separately. The current implementation uses
``MinMax'' if its region is smaller than twice the number of non-zero values in this row
and ``Sort'' in all other cases. In Figure~\ref{fig:minmax_stdsort_random} the switch
from ``MinMax'' to ``Sort'' is clearly visible between $N=49$ and $N=64$. We found that
as long as the storing method is not about to change, the ``Combined'' kernel is at
most 5\% less efficient than the kernels with only a single strategy. Overall, the ``Combined''
kernel reaches 35\% of the pure computation performance for the \mbox{CSR $\times$ CSR}
test case using the five-point finite difference stencils.

\begin{figure}[h]
   \centering
   \includegraphics*[width=0.48\textwidth]{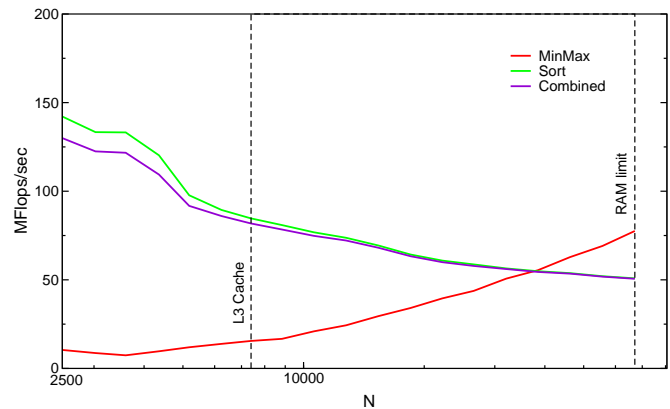}
   \caption[Comparison of the ``MinMax'' approach with the ``Sort'' approach multiplying randomly generated matrices with a fixed 0.1\% fill-rate]{
      \label{fig:minmax_stdsort_0_001}
      Comparison of the ``MinMax'' approach with the ``Sort'' approach, multiplying randomly generated matrices with a fixed 0.1\% fill ratio.
   }
\end{figure}

All previously shown test-cases used matrices with a fixed number of non-zero entries in each row.
This means that the fill ratio decreases with increasing problem size. The benchmark shown in 
Figure~\ref{fig:minmax_stdsort_0_001} uses the same matrix generation algorithm as 
for the random case, but the fill ratio is 0.1\% for each row instead of the fixed five elements. 
With the increasing absolute number of non-zero entries in each row the fill ratio of the result 
matrix increases. At $N\approx 38000$ the ``MinMax'' approach exceeds the performance of the ``Combined''
kernel, which uses the ``Sort'' storing strategy. At this point the fill ratio of the result 
matrix is 3.7\% or about 1400 non-zero entries per row. For the ``MinMax'' kernel this means that
on average every third cache line loaded actually contains one non-zero entry. Our conclusion
is that there is a break-even point in terms of problem size for which the ``MinMax'' approach 
is faster than sorting because of the growing probability that loaded data is actually 
stored in the result matrix.

 

\section{Performance Comparison of SET Libraries}
\label{sec:results}

In this section we compare the performance of the \blaze{} library to other expression template based C++ libraries.
We selected the most common libraries that provide the according kernels for the multiplication of two CSR matrices
and the multiplication of a CSR and a CSC matrix. We use the Boost \ublas{} library in version 1.51, \mtl{} in version
4.0.8883 (open source edition), \eigen{} in version 3.1.1, and \blaze{} in version 1.1, the latter employing the fastest
``Combined'' kernel from Section~\ref{sec:storing}. All libraries were benchmarked as given. We only present
double precision results in MFlop/s graphs for each test case. For all in-cache benchmarks we make sure that the
data has already been loaded to the cache.

\begin{figure}[h]
   \centering
   \includegraphics*[width=0.48\textwidth]{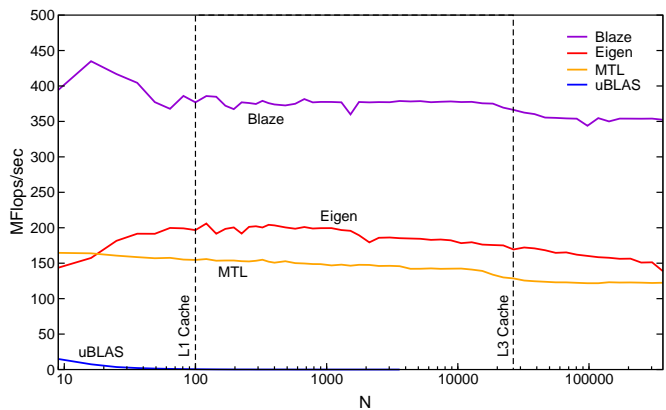}
   \caption[Results for the CSR = CSR $\times$ CSR benchmark (FD)]{
      \label{fig:csr_csr_fd_results}
      Performance comparison for the CSR = CSR $\times$ CSR benchmark (FD).
   }
\end{figure}

Figure~\ref{fig:csr_csr_fd_results} shows the comparison of the results of the \mbox{CSR $\times$ CSR} kernels
for sparse matrices resulting from five-point finite difference stencils. The \blaze{} library achieves roughly twice the
performance of \eigen{} and \mtl{}. \ublas{} cannot compete with the others, since it abstracts
from the actual storage order of the operands and traverses the right-hand side operand in a column-wise fashion
despite it being stored in row-major order. It becomes apparent that with a proper implementation of the kernel the
size of the matrix hardly influences the performance. Only a small drop  can be observed for matrices
that do not fit into the L3 cache anymore and have to be loaded from main memory.

\begin{figure}[h]
   \centering
   \includegraphics*[width=0.48\textwidth]{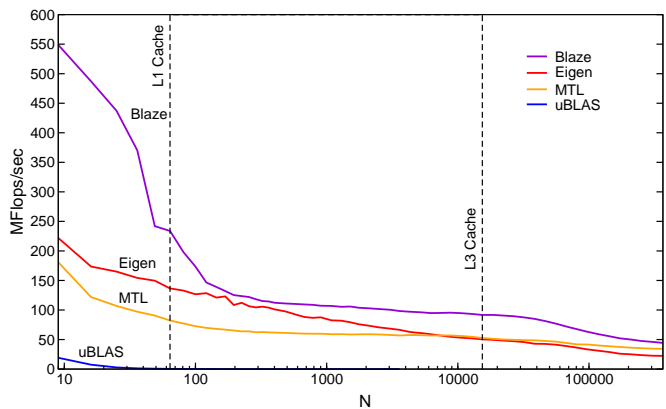}
   \caption[Performance comparison for the CSR = CSR $\times$ CSR benchmark (random)]{
      \label{fig:csr_csr_rnd_results}
      Performance comparison for the CSR = CSR $\times$ CSR benchmark (random).
   }
\end{figure}

Figure~\ref{fig:csr_csr_rnd_results} summarizes the results for the \mbox{CSR $\times$ CSR} kernels for randomly
created sparse matrices. Again, \blaze{} shows a higher performance than the \eigen{} and \mtl{} libraries, and \ublas{}
falls far behind. In comparison to sparse matrices resulting from finite difference stencils, though, the performance
clearly depends on the size of the matrix and degrades with growing matrix sizes.

\begin{figure}[h]
   \centering
   \includegraphics*[width=0.48\textwidth]{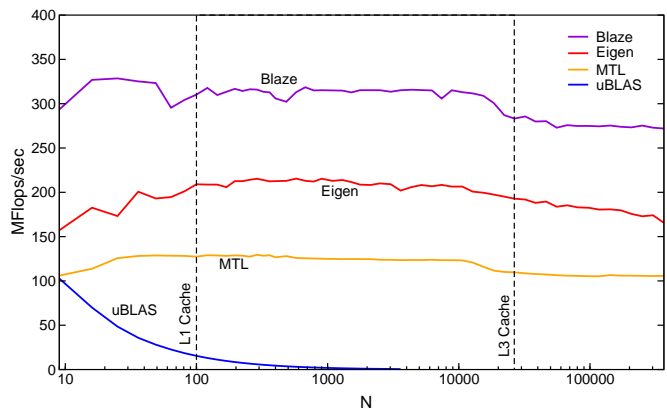}
   \caption[Results for the CSR = CSR $\times$ CSC benchmark (FD)]{
      \label{fig:csr_csc_fd_results}
      Performance comparison for the CSR = CSR $\times$ CSC benchmark (FD).
   }
\end{figure}

The results for the \mbox{CSR $\times$ CSC} kernels for sparse matrices resulting from five-point finite difference
stencils are presented in Figure~\ref{fig:csr_csc_fd_results}. The performance of the \blaze{} and \mtl{} libraries
drop due to the creation of a temporary CSR matrix and converting the storage order of the right-hand side
operand. The performance of \eigen{} slightly increases in comparison to the \mbox{CSR $\times$ CSR} kernel.
Also the performance of the \ublas{} library increases since the strategy of multiplying a row and a column
fits the given storage orders. However, still the performance drops quickly with growing problem size and 
prohibits the multiplication of large sparse matrices.

\begin{figure}[h]
   \centering
   \includegraphics*[width=0.48\textwidth]{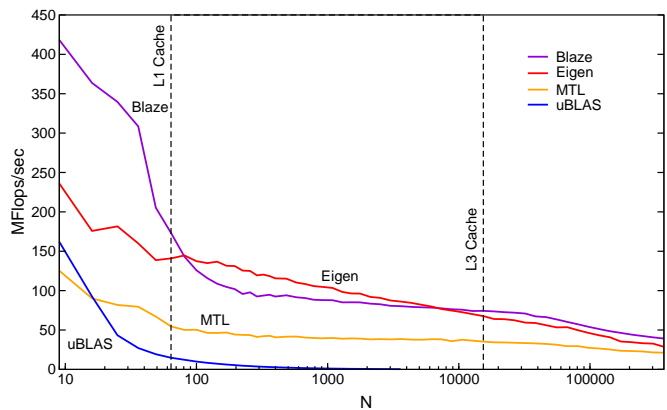}
   \caption[Performance comparison for the CSR = CSR $\times$ CSC benchmark (random)]{
      \label{fig:csr_csc_rnd_results}
      Performance comparison for the CSR = CSR $\times$ CSC benchmark (random).
   }
\end{figure}

Finally, Figure~\ref{fig:csr_csc_rnd_results} shows the results for \mbox{CSR $\times$ CSC} kernels for random sparse
matrices. Again, the performance of the \blaze{} and \mtl{} libraries drop to the creation of a converted temporary
and the performance of \eigen{} slightly increases. Consequently,  the performance of
\eigen{} can even surpass the \blaze{} performance for medium-sized matrices. 
For small and large sparse matrices \blaze{} exhibits
the best performance.

\section{Conclusion and Future Work}
\label{sec:conclusion}

We have conducted the first thorough performance analysis of
several spMMM kernels on a modern standard processor.
Employing a simple performance model we have demonstrated that our 
implementations can come close to
the maximum predicted performance in the computational part of the
kernel for out-of-cache situations with matrices leading to 
streaming memory access patterns. Due to further optimizations in the
memory management and storage strategy, we can provide the currently
fastest C++-based spMMM as part of the \blaze{} C++ library. \blaze{} combines
high maintainability, which proves to be of essential importance for
large scale software development, with HPC-grade performance that
matches or exceeds the capabilities of other commonly used C++ math 
libraries.

With the single core performance optimized the next step to improve
the \blaze{} library is to include shared memory parallelization to
exploit many- and multicore architectures. We expect that the typical
contention and saturation effects seen with these architectures will
add many new effects to the results presented here. Additionally, more work has
to be invested in further improving the single core performance.
Exploiting the given structure of the sparse matrix operands might be
a possible approach. Alternative sorting algorithms which 
are better suited to sort short lists of unique integral numbers may also
be advantageous. Finally,
the decision criterion for which of the two storing strategies to use
might be further improved.





%

\bibliographystyle{IEEEtran}
\bibliography{literature}

\end{document}